\documentclass[a4paper,11pt]{article}
\usepackage{pos}
\usepackage{bbm}

\title{Chiral Transition via Strong Coupling Expansion}
\author[a]{Jangho Kim}
\author*[b]{Pratitee Pattanaik}
\author[b]{Wolfgang Unger}

\affiliation[a]{Institute for Advanced Simulation (IAS-4), Forschungszentrum J\"ulich,\\
Wilhelm-Johnen-Straße, 52428 J\"ulich, Germany}

\affiliation[b]{Fakult\"at f\"ur Physik, Universit\"at Bielefeld,\\
Universit\"atstrasse 25, D33619 Bielefeld, Germany}

\emailAdd{j.kim@fz-juelich.de}
\emailAdd{pratiteep@physik.uni-bielefeld.de}
\emailAdd{wunger@physik.uni-bielefeld.de}

\newcommand{\ijkl}{_{i\,^{j},k\,^{l}}}
\newcommand{\Nc}{N_c}
\newcommand{\Tr}{{\rm Tr}}
\newcommand{\Ord}{{\mathcal{O}}}

\abstract{We investigate the chiral transition of $U(3)$ lattice gauge theory based on the strong coupling expansion. A generalized vertex model with vertices and weights derived from the tensor network approach of the dual representation of lattice QCD with staggered fermions is used and the configurations are sampled by the Metropolis algorithm. We study the chiral transition in the chiral limit and focus on the dependence of the second-order chiral transition temperature $aT_c$ for different values of the lattice gauge coupling $\beta$. We compare different orders of truncations of the strong coupling expansion: $\Ord(\beta^0)$, $\Ord(\beta^1)$, and $\Ord(\beta^2)$. We comment on the prospects of extending to $SU(3)$ at finite density.}

\FullConference{%
  40th International Symposium on Lattice Field Theory-Lattice2023,\\
  31st August - 4th September, 2023\\
  FNAL, USA
}

\begin{document}
\maketitle

\section{Introduction}

We have a limited understanding of the $\mu_B-T$ phase diagram of quantum chromodynamics. Lattice QCD (LQCD) gives us access to the $\mu_B/T<3$ region of the phase diagram while beyond this region, the phase diagram is inaccessible due to the infamous sign problem. We know that at low $\mu_B$, there is a crossover chiral transition between hadrons and quark-gluon plasma. It is generally believed that this crossover transition ends in a critical end point (CEP) and is followed by a first-order transition as $\mu_B$ increases. To find this unknown location of CEP and the existence of the first-order phase transition, alternative approaches are used to overcome the sign problem. One of these approaches is to study LQCD using the strong coupling expansion.\\
In this approach, the order of integration is reversed i.e. the gauge integration is performed first, followed by the Grassmann integration. 
The strong coupling limit $\beta=0$ was first studied via meanfield theory \cite{Kawamoto1981}, and via Monte Carlo by using the so-called dual formulation of LQCD\cite{Rossi1984}. The dual formulation is well established in the strong coupling limit and the complete $\mu_B-T$ phase diagram is determined \cite{Karsch1989,Forcrand2010}. At leading order $\Ord(\beta)$, the sign problem is still mild enough to study the phase diagram \cite{deForcrand2014} and was recently used to determine the nuclear liquid-gas transition \cite{Kim2023}. For higher orders of $\beta$, $\Ord(\beta^n)$, a promising approach was developed where the dual formulation results in a tensor network \cite{Gagliardi2019}. 
In this study, we have mapped the tensor network to a vertex model suitable for Monte Carlo simulations (in contrast to the tensor-network methods developed for SC-LQCD in \cite{Bloch2022}). We discuss the validity range in $\beta$ by comparing different orders of $\beta$. We also study the $\beta$-dependence of the chiral CEP in the chiral limit.

\section{Strong Coupling Expansion} 

The partition function of LQCD is given by
\begin{align}
Z=\int_G DU_{n, \mu} d\bar{\chi}(n) d\chi(n) \exp \left(-S_F[\bar{\chi}, \chi, U]\right) \exp \left(-S_G[U]\right)
\end{align}
where $DU$ is the Haar measure, $U\in SU(3)$ are the gauge fields on the lattice links $(x,\mu)$ and {$\bar{\chi_n},\chi_n$} are the unrooted staggered fermions at the lattice sites $n$. The gauge action $S_G[U]$ is given by the Wilson plaquette action:
\begin{align}
 S_G[U]=-\frac{\beta}{2N_c} \sum_p \operatorname{Re} \left(\operatorname{tr}\left[U_p\right]+\operatorname{tr}\left[U_p^{\dagger}\right]\right)
\end{align}
and $\beta$ is the inverse gauge coupling, $N_c$ is the number of colors and the summation runs over all plaquettes $p$. For the fermionic action $S_F[\bar{\chi}, \chi, U]$, we use the staggered fermionic discretization with the usual staggered phases, $\eta_\mu(n)$ and quark mass $m_q$ to get:
\begin{align}
 S_F[\bar{\chi}, \chi, U]=\sum_n\left(-\sum_\mu \eta_\mu(n)\left(\bar{\chi}_n U_{\hat{\mu}}(n) \chi_{n+\hat{\mu}}-\bar{\chi}_{n+\hat{\mu}} U_{\hat{\mu}}^{\dagger}(n) \chi_{n}\right)+2m_q\bar{\chi}\chi\right)
\end{align}
In the strong coupling regime, $\beta$ is small, and the gauge action can be Taylor-expanded in $\beta$:
\begin{align}
e^{-S_G[U]}=\sum_{n_p,\bar{n}_p=0}^{\infty} \frac{1}{n_p!} \frac{1}{\bar{n}_p!}\left(\frac{\beta}{2 \Nc}\right)^{n_p+\bar{n}_p}\left(\sum_p\operatorname{tr}U_p\right)^{n_p}\left(\sum_P \operatorname{tr}U_p^{\dagger}\right)^{\bar{n}_p},
\end{align}
which yields the exponents $n_p$ and $\bar{n}_p$ of the elementary (anti-) plaquette ($\operatorname{tr}U_p^{\dagger}$) $\operatorname{tr}U_p$, called the (anti-) plaquette occupation numbers. Taylor expansion in quark mass $m_q$ yields exponents $m_n$ called the monomer occupation number. These denote the number of color singlet mesonic states living at a site $n$. The massless Dirac operator can also be expanded in forward and backward directions to get dual variables $\{d_l, \bar{d}_l\}$, the minimum $k_l=\min(d_l,\bar{d}_l)$ called dimers which are meson hoppings on a link, and the unpaired (anti-)quark flux is given by the difference $f_l=d_l-\bar{d}_l$. The Partition function in terms of the dual variables becomes:
\begin{align}
Z\left(\beta, m_q\right)=\sum_{\substack{\left\{n_p, \bar{n}_p\right\} \\\left\{d_l, \bar{d}_l, m_n\right\}}} \prod_p \frac{\tilde{\beta}^{n_p+\bar{n}_p}}{n_{p}!\bar{n}_p!} \prod_l \frac{1}{d_{l}!\bar{d}_l!} \prod_x \frac{(2m_q)^{m_n}}{m_n!} \mathcal{G}_{n_p,\bar{n}_p,d_l,\bar{d}_l,n_x}
\end{align}
with $\mathcal{G}$ denoting the non-local gauge and Grassmann integral over the whole lattice. The non-local integral can be obtained by writing it in terms of generalized Weingarten functions and integer partitions. This allows for a different parametrization of the integral using irreducible matrix elements of the symmetric group. With this the integral decouples into decoupling operators $P^{\rho}$ with $\rho$ being the decoupling operator index (DOI): 
\begin{align}
\mathcal{I}^{a,b}\ijkl &=  \int_{SU(\Nc)}DU\,\, U_{i_{1}}^{\,j_{1}} \cdots U_{i_{a}}^{\,j_{a}}U^{\dag \,l_{1}}_{k_{1}} \cdots U^{\dag \,l_{b}}_{k_{b}}=\sum_{\rho}\left(P^{\rho}\right)^{\;l}_{i}\left(P^{\rho}\right)^{\;j}_{k},
\end{align}
where for a non-zero integral $|a-b| \mod \Nc =0 $, and we make use of the multi-index notation:
\begin{align}
i&=(i_1, \ldots, i_a),& 
j&=(j_1, \ldots, j_a),&
k&=(k_1, \ldots, k_b),&
l&=(l_1, \ldots, l_b),
\end{align} and $\rho$ are multi-indices that are mapped to a set of positive integers \cite{Gagliardi2019}. 
Fixing the DOI lets us contract color indices independently at each site. This contraction of color indices depends on the dual variables {$n_p,\bar{n}_p,d_l,\bar{d}_l$}. There are two types of color indices in $\mathcal{G}_{n_p,\bar{n}_p,d_l,\bar{d}_l,m_n}$: The first type is fermionic and arises from the expansion of hopping terms {$d_l,\bar{d}_l$} which are contracted with colored Grassmann fields. The second type is gluonic and is due to the expansion of the Wilson action, they are contracted according to the plaquettes governed by {$n_p,\bar{n}_p$}. The integration and the contraction of the color indices results in a tensor that depends on the DOI $\{\rho^{n}_{d}\}$ for each direction:
\begin{align}
T_{n}^{\rho^{n}_{-d}\cdots\rho^{n}_{d}}\left(\mathcal{D}_{x}\right)&\equiv \Tr_{\mathcal{D}_{n}}\left[{\displaystyle \prod_{\pm \mu}}P^{\rho^{n}_{\mu}}\right]\in\mathbbm{R},& 
\mathcal{D}_{n}&= \left\{m_n,d_{n,\pm\mu}, n_{n,\mu\nu},\bar{n}_{n,\mu\nu}\right\}
\end{align}
and with this, the partition function only depends on local variables:
\begin{align}
&Z\left(\beta, \mu_q, m_q\right)=\nonumber\\
&\sum_{\substack{\left\{n_p, \bar{n}_p\right\} \\\left\{k_{\ell}, f_{\ell}, m_n\right\}}} \sigma_f \sum_{\left\{\rho_\mu^n\right\}} \prod_p \frac{\left(\frac{\beta}{2 N}\right)^{n_p+\bar{n}_p}}{n_{p} ! \bar{n}_{p} !} \prod_{\ell=(n, \mu)} \frac{e^{\mu_q \delta_{\mu, 0} f_{n, \mu}}}{k_{\ell} !\left(k_{\ell}+\left|f_{\ell}\right|\right) !} \prod_n \frac{\left(2 m_q\right)^{m_n}}{m_{n} !} T_n^{\rho_{-d}^n \ldots \rho_{+d}^n}\left(\mathcal{D}_n\right)
\end{align}
as only the local dual variables $\mathcal{D}_n$ enter the tensor. 
The fermionic sign is denoted by $\sigma_f$, which includes the staggered phases and the global geometric sign due to the quark fluxes. The tensors can also have negative real values, resulting in a negative tensor sign $\sigma_t$. The total sign is the product of both: $\sigma=\sigma_t \sigma_f$. For valid configurations on the lattice, two constraints must be fulfilled. The first is the Grassmann constraint which requires that exactly $N_c$ fermion and $N_c$ anti-fermion fields must be present at a site: 
\begin{align}
m_n+\sum_{\pm \mu} \left( k_{n,\mu} +\frac{|f_{n,\mu}|}{2} \right)=N_c \quad \quad  \sum_{\pm \mu} f_{n,\mu}=0
\end{align}
The second constraint is the gauge constraint with the requirement that the sum of gauge flux and fermion flux on each link of the lattice should be an integer multiple of $N_c$:
\begin{align}
f_{n,\mu}+\sum_{\nu>\mu}\left[ \delta n_{\mu,\nu}(n)-\delta n_{\mu,\nu}(n-\nu)\right] =N_c\mathbbm{Z}
\end{align}
This summarizes the tensor network representation of LQCD, in principle valid to all orders in $\beta$.

\section{From tensor network representation to vertex model}

While LQCD in the strong coupling limit could be efficiently simulated via the Worm algorithm, this is no longer the case in the more general tensor network representation. For an alternative Monte Carlo algorithm, we can map the set of tensors on a vertex model (see \cite{Wenger2008} for a similar formulation for Wilson fermions at strong coupling in 2 dimensions). For simplicity, we here consider the $U(3)$ gauge theory in 4 dimensions, up to $\Ord(\beta^2)$ gauge corrections. In contrast to SU(3) (to be presented in a forthcoming publication), the gauge group U(3) has $\sigma_f=1$, thus the sign problem only occurs in $\sigma_t$. The key step in the calculation of vertices from the tensors is to define the link state as an integer from a combination of DOI $\rho_l$, $d_l$ and $\bar{d}_l$, and the $n_p$, $\bar{n}_p$ adjacent to that link. The computation has been automatized for a given order in $\beta$, the gauge group $SU(\Nc)$ or $U(\Nc)$, dimension $d$, and results in a list of valid vertices and the corresponding weights. Monte Carlo is in terms of these vertices, but given a vertex configuration, the dual degrees of freedom can be parsed to obtain observables. An example of vertices for the gauge group U(3) with $\Ord(\beta^2)$ in $1+1$ dimension is shown in Fig.~1. A valid configuration on the lattice is one where adjacent vertices have the same value on the edge shared between them. A sample configuration with the vertices is shown in Fig.~2(a). In Fig.~2(b), the same configuration is shown explicitly with dual variables.
We sample the vertex configurations via the Metropolis algorithm with the weight of a vertex calculated by the product of the tensor weight $w_T=T_n^{\rho^{n}_{-d}\cdots\rho^{n}_{d}}$ and the weight due to the parameters of the theory: quark mass $am_q$, 
bare anisotropy $\gamma$ (which is related via the physical anisotropy $\xi$ to the temperature $aT=\xi(\gamma)/N_t$), inverse gauge coupling $\beta$, and baryon chemical potential $a\mu_B$: 
\begin{align}
w_D(am_q,\gamma,\beta,a\mu_B)= (2am_q)^{m_n} \gamma^{\frac{1}{2}(k_{n,t\pm}+f_{n,t+}+f_{n,t-})}\left(\frac{\beta}{2\Nc}\right)^{\frac{1}{4}(n_p+\bar{n}_p)} (e^{a\mu_B/\gamma^2})^{\frac{1}{2}(f_{n,t+}-f_{n,t-})} %w_D(am_q,\gamma,\beta,a\mu_B)= (2am_q)^{N_m} \gamma^{\frac{1}{2}(N_{D_t}+N_{F_{t,+}}+N_{F_{t,-}})}\left(\frac{\beta}{2\Nc}\right)^{\frac{1}{4}(N_p+\bar{N}_p)} (e^{a\mu_B/\gamma^2})^{\frac{1}{2}(N_{F_{t,+}}-N_{F_{t,-}})}
\end{align}
which is computed based on the dual variables and assigned to the vertex weight: $w_V=w_T w_D$. Here, the subscript $t\pm$, $t+$, and $t-$ of the dual variables $k$ and $f$ denote the positive/negative temporal direction, the positive temporal direction, and the negative temporal direction of the vertex respectively. 
The acceptance probability is the usual Metropolis acceptance probability:
\begin{align}
\min \left(1,\frac{\prod_{i\in\mathcal{C}} w(v_i^{\rm{new}})}{\prod_{i\in\mathcal{C}} w(v_i^{\rm{old}})}\right)
\end{align}
where the product is over all vertices that are to be modified during one update, with $\mathcal{C}$ a closed loop that is either a $1\times1$  loop or a line update in the spatial or temporal direction. 
The first is called a plaquette update (as they can change $n_p$, $\bar{n}_p$) as shown in Fig.~3a, the second is called a line update as shown in Fig.~3b. Mixing both updates gives an ergodic algorithm. 
We have parallelized the code to perform updates on independent contours simultaneously. 

\begin{figure}[htbp]
    \centering
    \begin{subfigure}[b]{0.6\textwidth}
        \centering
        \includegraphics[width=\textwidth]{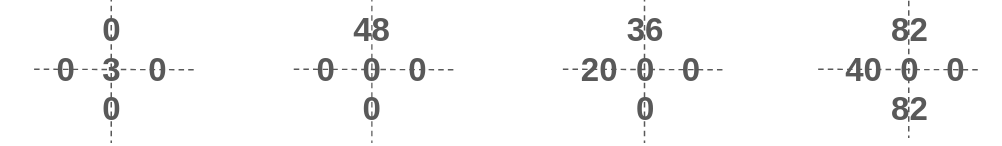}
        \caption{Vertices used in our simulations. Each vertex has a weight which is not shown here.}
        \label{Fig1a}
    \end{subfigure}
    \begin{subfigure}[b]{0.6\textwidth}
        \centering
        \includegraphics[width=\textwidth]{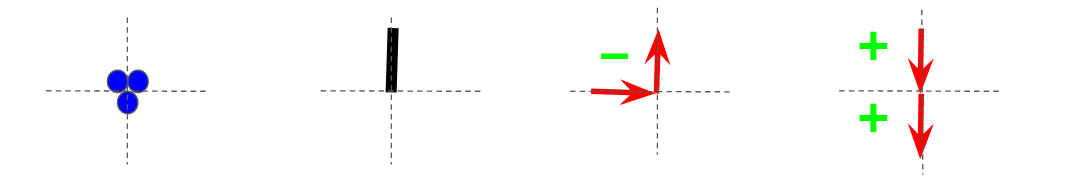}
        \caption{Same vertices as above represented by their dual variables content. The blue circle denotes a monomer, the black line denotes a dimer, the red arrow denotes a quark flux, and the plaquettes are denoted by $+$ or $-$ depending on their orientation.}
        \label{Fig1b}
    \end{subfigure}
    \caption{A few valid vertices of gauge group $U(3)$ with $\Ord(\beta^2)$ gauge corrections in $1+1$ dimension.}
    \label{Fig1}
\end{figure} 

\begin{figure}[htbp]
    \centering
    \begin{subfigure}[b]{0.35\textwidth}
        \centering
        \includegraphics[width=\textwidth]{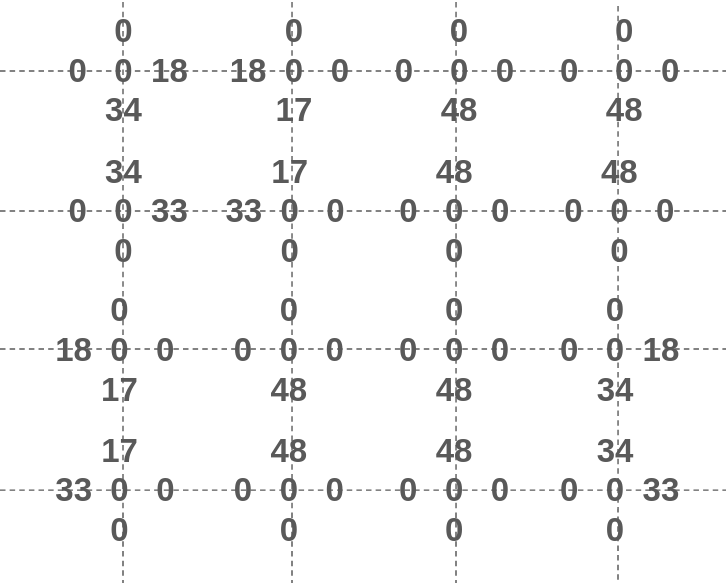}
    \end{subfigure}
    \begin{subfigure}[b]{0.35\textwidth}
        \centering
        \includegraphics[width=\textwidth]{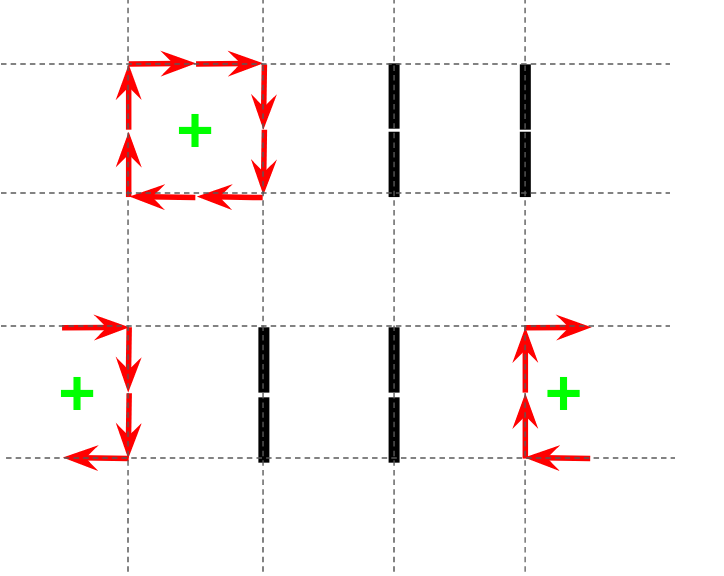}
    \end{subfigure}
    \caption{A valid configuration of gauge group $U(3)$ on $4\times4$ periodic lattice with $\Ord(\beta^2)$. \textit{Left}: Configuration shown with vertices. \textit{Right}: The same configuration is shown with dual variables for better understanding.}
    \label{Fig2}
\end{figure} 

\begin{figure}[htbp]
    \centering
    \begin{subfigure}[b]{0.7\textwidth}
        \centering
        \includegraphics[width=\textwidth]{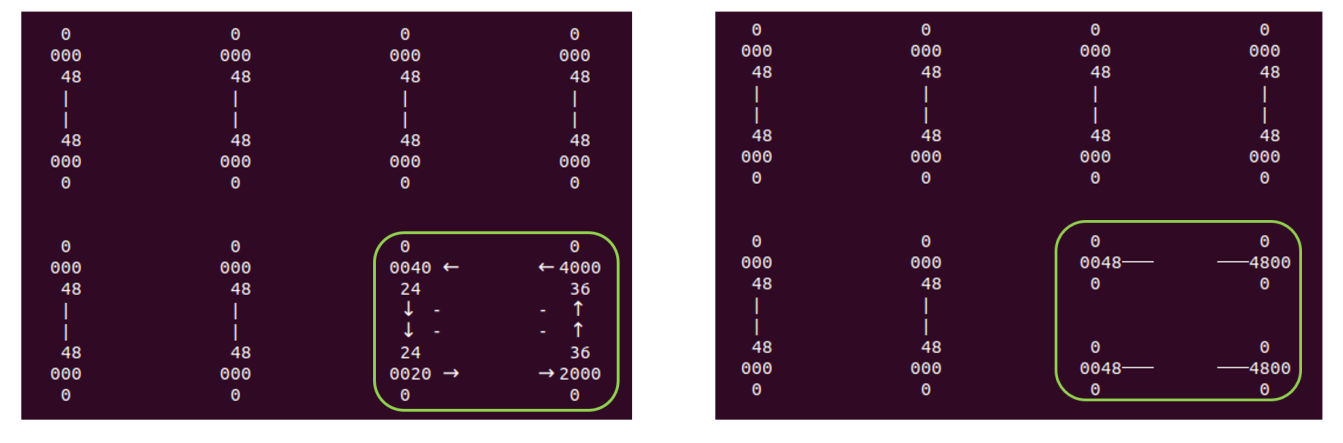}
        \caption{Example of a plaquette update where the second configuration is obtained after changing $4$ vertices.}
        \label{Fig3a}
    \end{subfigure}
    \\
    \begin{subfigure}[b]{0.7\textwidth}
        \centering
        \includegraphics[width=\textwidth]{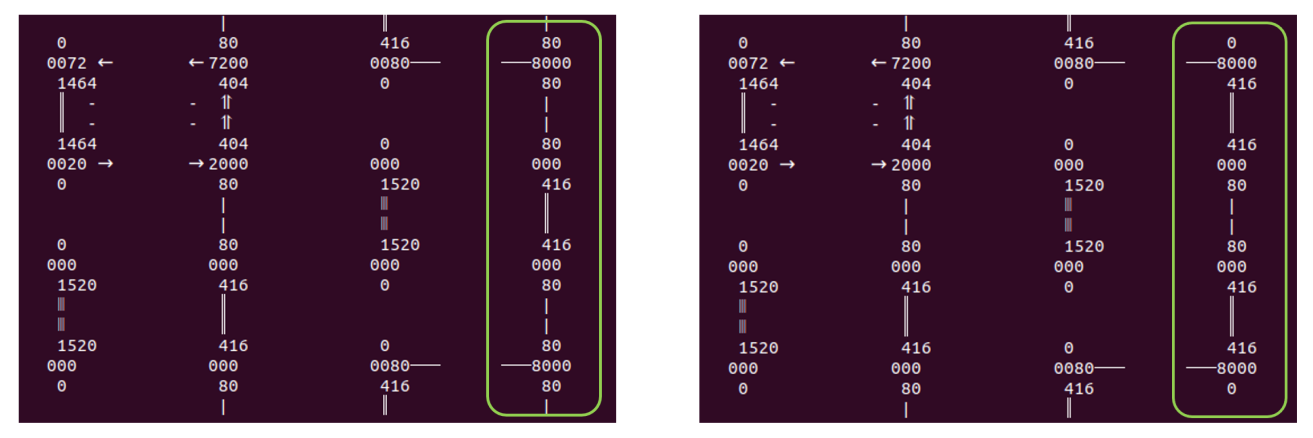}
        \caption{Example of a line update where the second configuration is obtained after changing all the vertices on a line.}
        \label{Fig3b}
    \end{subfigure}
    \caption{Updates required to sample all possible configurations of gauge group $U(N)$. Here, the configurations are shown in a hybrid format with vertices as well as dual variables for better understanding. In actual updates, only the vertices are required.}
    \label{Fig3}
\end{figure}

\section{Validation of the model and comparison between different orders of $\beta$} 
The observables we consider to determine the second-order transition temperature and its $\beta$ dependence are energy $\langle E\rangle$, specific heat $C_v$, and average plaquette $\langle P\rangle$, which in terms of dual variables are:
\begin{align}
\langle E\rangle=\frac{\langle N_{D_t}+N_{F_{t,+}}+N_{F_{t,-}} \rangle}{V} \quad \quad C_v =\frac{\langle E^2\rangle-\langle E \rangle^2}{V} \quad \quad
\langle P\rangle=\frac{\langle N_p+ \bar{N}_p \rangle}{2\beta V}
\end{align}
Here, $N_{Dt}$ denotes the total number of temporal dimers, $N_{Ft,\pm}$ denotes the total number of temporal fermion flux in the positive/negative direction, $N_p, \bar{N}_p$ denote the total numbers of (anti-) plaquettes, in a configuration. 
Since we are working in the chiral limit, we cannot measure the chiral condensate. The chiral limit pushes the crossover transition to a second-order transition. The specific heat peaks at the transition temperature which allows us to determine the second-order transition temperature. 
To verify that the Metropolis sampling with the vertex model yields the correct result, we compare the Monte Carlo data with exact enumeration on $2\times 2$ lattice, showing perfect agreement: Fig.~4a shows the specific heat as a function of $\beta$ for two different temperatures. $\Ord(\beta^0)$ Metropolis data with the vertex model is shown in black, $\Ord(\beta^1)$ data is shown in green, and $\Ord(\beta^2)$ is shown in red. The dotted lines denote the exact enumeration results. Fig. 4b. shows the specific heat as a function of temperature for $\Ord(\beta^0)$, $\Ord(\beta^1)$, and $\Ord(\beta^2)$ for two different $\beta$ values. Fig. 4c. shows the average plaquette as a function of $\beta$ for different orders in $\beta$ and temperatures. Clearly, this gauge observables has the strongest dependence on the order, implying that including $\Ord(\beta^2)$ is important.
\begin{figure}[htbp]
    \centering
    \hspace{-2cm}
    \begin{subfigure}[b]{0.35\textwidth}
        \centering
        \includegraphics[width=\textwidth]{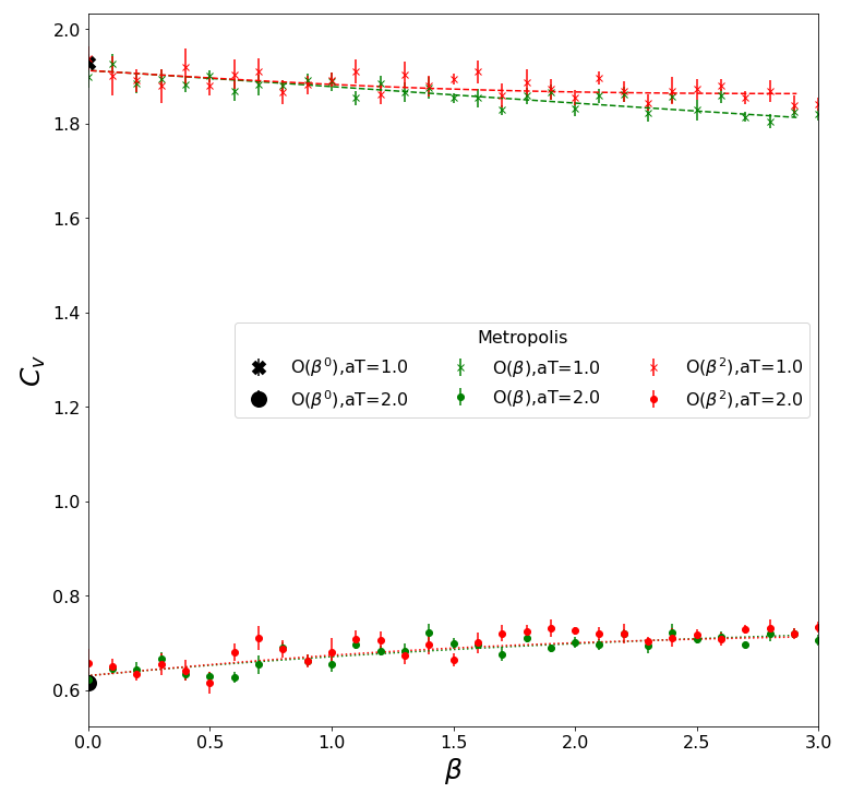}
        \caption{Specific heat as a function of $\beta$.}
        \label{Fig4a}
    \end{subfigure}
    \begin{subfigure}[b]{0.35\textwidth}
        \centering
        \includegraphics[width=\textwidth]{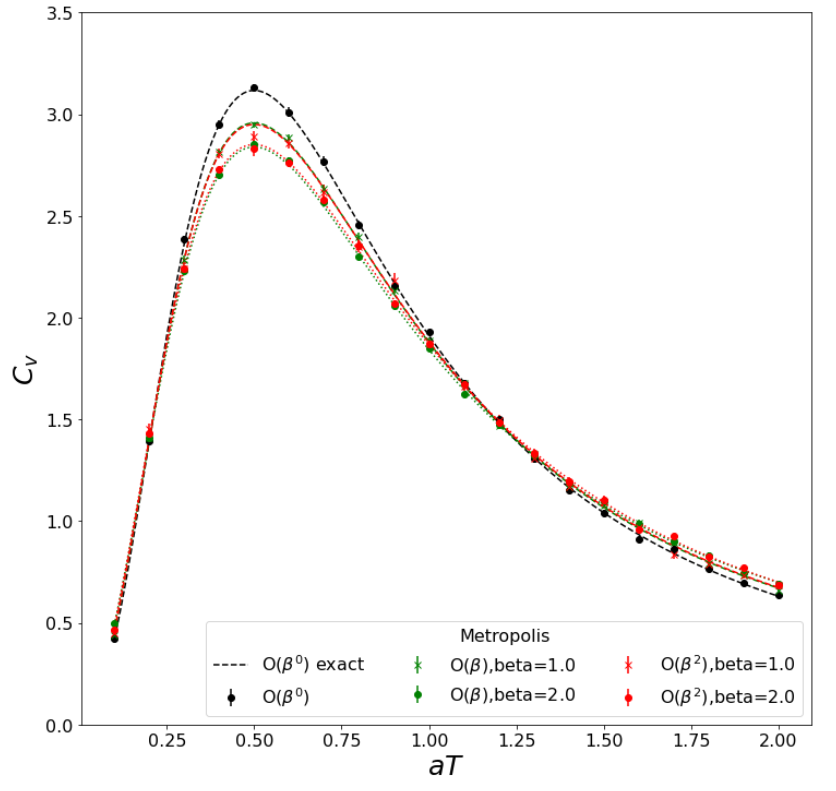}
        \caption{Specific heat as a function of $aT$.}
        \label{Fig4b}
    \end{subfigure}
    \begin{subfigure}[b]{0.35\textwidth}
        \centering
        \includegraphics[width=\textwidth]{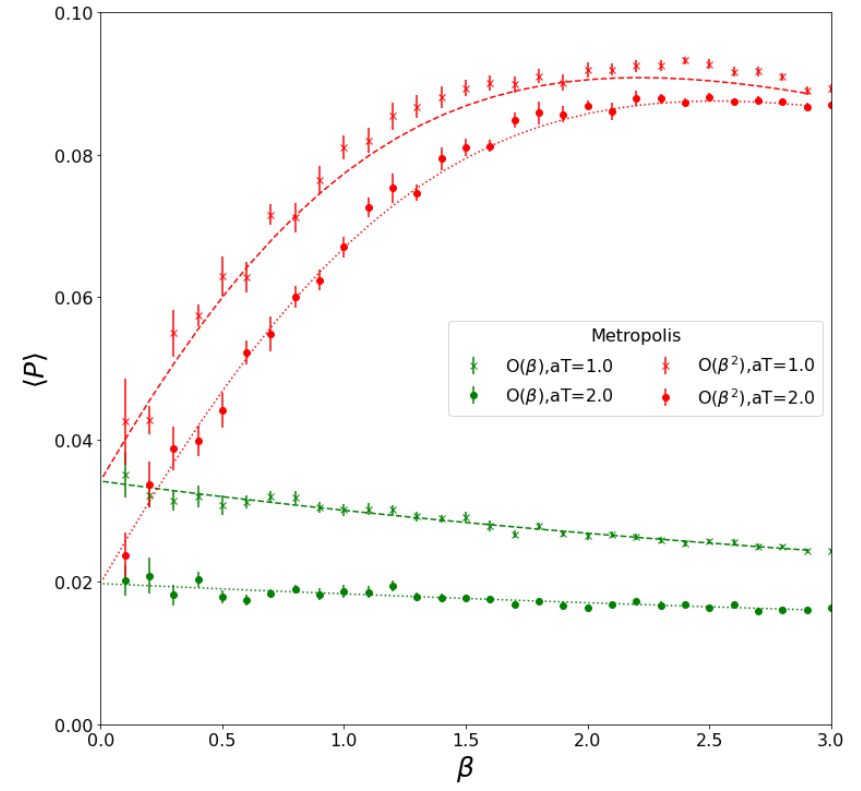}
        \caption{$\langle P \rangle$ as a function of $\beta$}
        \label{Fig4c}
    \end{subfigure}
    \hspace{-2cm}
    \caption{The results are for $2\times 2$ lattice of gauge group $U(3)$. The exact enumeration results are shown as solid lines for $\Ord(\beta)$ and dashed lines for $\Ord(\beta^2)$ and the points and crosses show the Metropolis data.}
    \label{Fig4}
\end{figure}

Fig.~5 shows the Metropolis results in a $8^3\times4$ lattice. We aim for larger volumes for a finite size scaling analysis in a forthcoming publication. At $\Ord(\beta)$, all configurations have positive weight, but $\Ord(\beta^2)$ introduces a sign problem due to tensor signs not being canceled. Fig.~5a shows $\Delta f=-\frac{T}{V}$ln$(\sigma)$ as a function of $aT$ for different $\beta$. Increasing $\beta$ makes the sign problem worse. Fig.~5b shows the specific heat for different $\beta$ as a function of $aT$. We model it with Pad\'e approximants which are defined as rational functions, $[L,M](\beta)=\frac{a_0+a_1\beta+\cdots+a_L\beta^L}{1+b_1\beta+\cdots+b_M\beta^M}$. The fit of the data with Pad\'e approximant [3,3], and the fitted peak position which indicates the crossover temperature $aT_c(\beta)$ is indicated with vertical lines. These results are summarized for both orders in $\beta$ in Fig.~4c as a function of $\beta$. We see that $aT_c$ decreases with increasing $\beta$ for $\Ord(\beta)$ which is expected because the lattice spacing decreases with increasing $\beta$. In contrast $aT_c$ for $\Ord(\beta^2)$ also decreases, however, it levels off, a behavior that is unexpected. We plan to perform simulations with smaller and larger values of $\beta$ to understand this further.

\begin{figure}[htbp]
    \centering
    \hspace{-2cm}
    \begin{subfigure}[b]{0.35\textwidth}
        \centering
        \includegraphics[width=\textwidth]{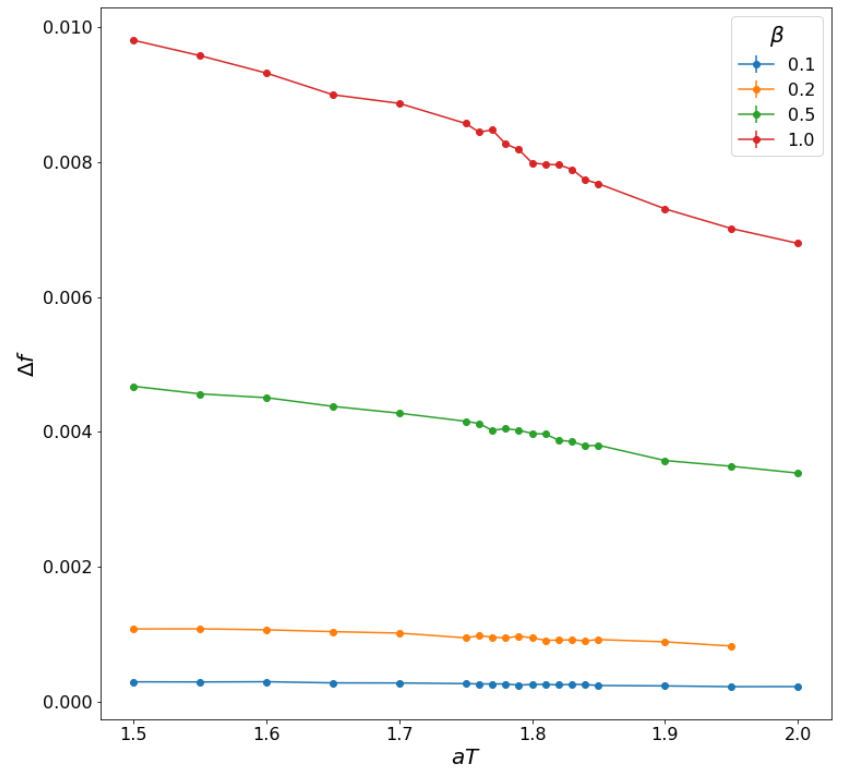}
        \caption{$\Delta f$ as a function of $aT$.}
        \label{Fig4c}
    \end{subfigure}
    \begin{subfigure}[b]{0.35\textwidth}
        \centering
        \includegraphics[width=\textwidth]{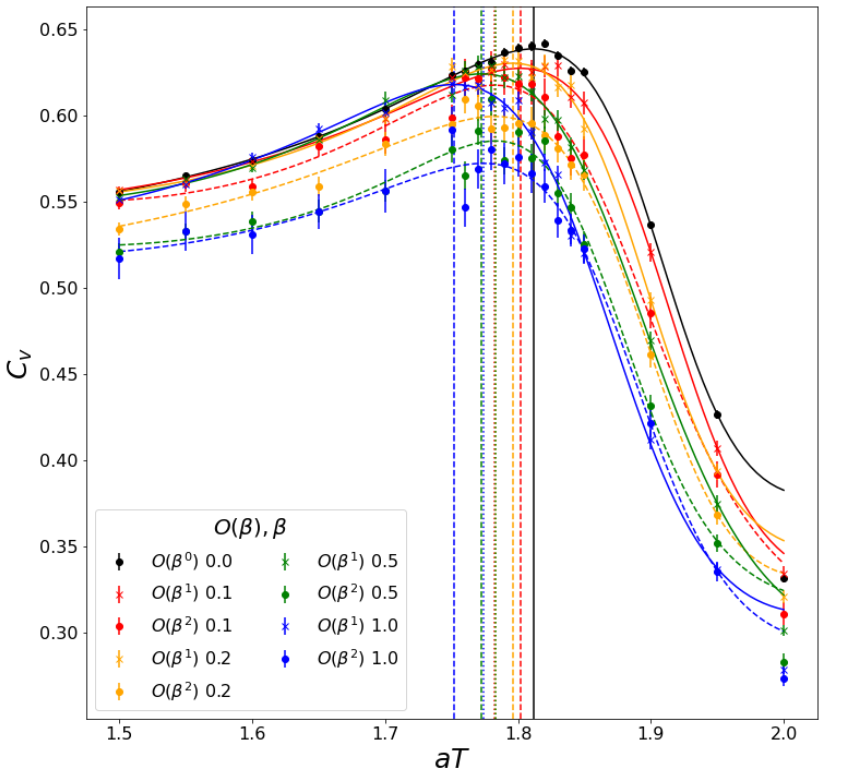}
        \caption{Specific heat as a function of $aT$.}
        \label{Fig4e}
    \end{subfigure}
    \begin{subfigure}[b]{0.35\textwidth}
        \centering
        \includegraphics[width=\textwidth]{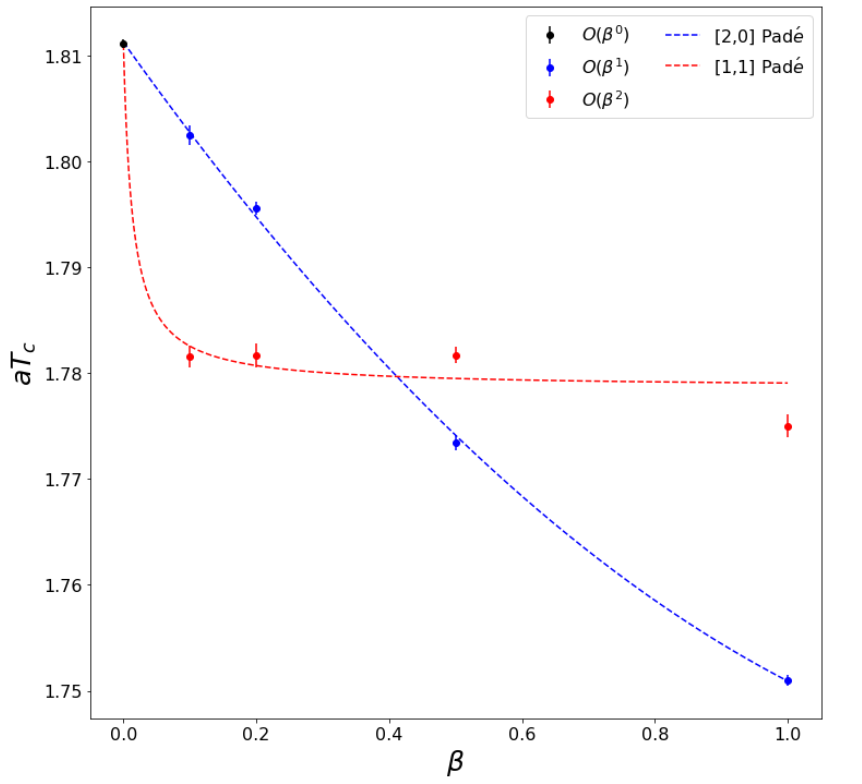}
        \caption{$aT_c$ as a function of $\beta$.}
        \label{Fig4f}
    \end{subfigure}
    \hspace{-2cm}
    \caption{All results shown are from Monte Carlo simulations on $8^3\times 4$ lattice of gauge group $U(3)$. \textit{Left}: $\Delta f$ is shown as a function of $aT$ for different values of $\beta$ with $O(\beta^2)$. Increasing $\beta$ makes the sign problem worse. \textit{Center}: The specific heat is shown as a function of $aT$ for different values and orders of $\beta$. The data is fitted with Pad\'e approximants $[3,3]$ and the peak position of the fit is shown with vertical lines which indicates $aT_c$. \textit{Right}: The second-order transition temperature $aT_c$ is shown as a function of $\beta$ for different orders of $\beta$. The data for $O(\beta)$ is fitted with Pad\'e approximants $[2,0]$ and the data for $O(\beta^2)$ is fitted with Pad\'e approximants $[1,1]$.}
    \label{Fig5}
\end{figure}

\section{Conclusion and Outlook}

The worm algorithm that has been successfully applied to the dual formulation of strong coupling, cannot be used when nontrivial decoupling operator indices are involved. Hence we came up with an alternative simulation strategy.
We simulated the vertex model obtained from tensor network representation for gauge group $U(\Nc)$ via the Metropolis algorithm for different orders of $\beta$. 
On small volumes, we could explicitly check the validity with exact enumeration. On a $8^3\times 4$ lattice, we determined the chiral transition in the chiral limit from the peak of the susceptibility as a function of  $\beta$ for different orders in $\beta$. We see that the transition temperature decreases as $\beta$ increases, although the $\Ord(\beta^2)$ dependence is unexpected. 
We have now also implemented a heatbath algorithm for the vertex model, improving on the performance. 
In the future, we plan to extend the model to gauge group $SU(\Nc)$ and perform simulations with non-zero quark baryon chemical potential for various $\beta$ in the vicinity of the chiral critical point located at $(a\mu_{B,c},aT_c)=(1.92(6),0.98(3))$ at strong coupling \cite{Klegrewe2020}. 

\section{Acknowledgments}
The authors gratefully acknowledge the funding of this project by computing time provided by the Paderborn Center for Parallel Computing ($PC^2$). This work is supported by the Deutsche Forschungsgemeinschaft (DFG) through the CRC-TR 211 ’Strong-interaction matter under extreme conditions’– project number 315477589 – TRR 211. J.K. was supported by
the Deutsche Forschungsgemeinschaft (DFG, German Research Foundation) through the funds provided to the Sino-German Collaborative Research Center TRR110 ”Symmetries and the Emergence of Structure in QCD” (DFG Project-ID 196253076 - TRR 110).

\end{document}